\documentclass[onecolumn]{aastex62}
\shorttitle{Delay}
\shortauthors{Penacchioni \& Civitarese}

\bibpunct{(}{)}{;}{a}{}{,}
\usepackage{savesym}
\savesymbol{iint}
\savesymbol{iiint}
\savesymbol{iiiint}
\savesymbol{idotsint}
\usepackage{amsmath}
\restoresymbol{AMS}{iint}
\restoresymbol{AMS}{iiint}
\restoresymbol{AMS}{iiiint}
\restoresymbol{AMS}{idotsint}
\usepackage{graphicx}			
\usepackage{physics}
\usepackage{xcolor}
\usepackage[utf8]{inputenc}

\begin{document}

\correspondingauthor{A.V. Penacchioni}
\email{ana.penacchioni@fisica.unlp.edu.ar}

\author{A.V. Penacchioni}
\affil{IFLP (CONICET), La Plata, Argentina}

\author{O. Civitarese}
\affil{IFLP (CONICET), La Plata, Argentina}
\affil{Department of Physics, University of La Plata (UNLP),
49 y 115 cc. 67, 1900 La Plata, Argentina}

\title{Time-delay between neutrinos and gamma-rays in short GRBs}

\begin{abstract}
In this work we make an estimate of the time-delay between signals, recorded at detectors on Earth, of neutrinos and photons originated in a short gamma-ray burst. We describe the geometry and dynamics of the system according to the Fireshell model. The delay in the photon's arrival time is produced because the system is originally opaque to radiation; thus, the photons remain trapped and thermalize until the transparency condition is reached. We calculate the time interval between neutrino- and photon- emission in the black hole frame and  transform it to the observer-frame using Lorentz transformations. We obtain a difference in the arrival time at Earth of $\Delta t_{Earth} \approx 854.57$ s.
\end{abstract}

\keywords{neutrinos - gamma-ray bursts: general - astroparticle physics}

\section{Introduction}
Recently, remarkable progress has been achieved concerning the detection of gravitational waves, high energy photons and neutrinos from cosmological sources, such as active galactic nuclei (AGNs) and gamma-ray bursts (GRBs). These detections in some cases seem to originate from the same source, as it was the case for the event which generated the gravitational waves GW 170817 detected by LIGO \cite{2017Natur.551...85A}, followed by gamma-ray emission detected by Fermi two seconds later \cite{2017ApJ...848L..14G} and X-ray emission detected by Chandra nine days later \cite{2017Natur.551...71T}, all coming from the same region in space. The signals turned out to be those of a short GRB from a double neutron star (NS-NS) merger. A similar event took place on September $22^{nd}$, 2017, when a high-energy neutrino event detected by IceCube was coincident in direction and time with a gamma-ray flare from the blazar TXS 0506+056 \cite{2017ATel10791....1T,2018Sci...361..147I,2018ApJ...863L..10A}.

Many GRB models predict neutrino-emission prior to or simultaneous with photon-emission. In this work we aim at the calculation of the time difference between the arrival of neutrinos and photons from a short GRB originated by a NS-NS merger. We have adopted the Fireshell model \cite{2009AIPC.1111..587B} to describe the dynamics of the source, as done in \cite{2019ApJ...872...73P}. According to this model, the inner part of the merged system collapses to a black hole (BH). An $e^{\pm}$ plasma is created during this process due to vacuum polarization. The plasma is in thermal equilibrium at temperature $T$. A fraction of these pairs annihilates to photons ($e^- - e^+ \rightarrow \gamma +\gamma$), and another fraction converts to neutrinos ($e^- - e^+ \rightarrow \nu_e + \overline{\nu_e}$). Neutrinos are weakly-interacting particles, then they will leave the system isotropically as soon as they are created, traveling at nearly the speed of light. The system is opaque to radiation, though. Radiation pressure makes the plasma expand as an optically thick and spherically symmetric fireshell at relativistic velocities, engulfing the baryonic matter left over in the process of gravitational collapse, still maintaining thermal equilibrium. Photons will remain trapped until the system becomes transparent. Only then they will be able to escape producing a flash of radiation called P-GRB. The Lorentz factor $\Gamma$ at transparency varies between $\Gamma \sim 100 - 1000$, depending both on the initial plasma energy and the amount of baryonic matter that was swiped during the relativistic expansion \cite{2000A&A...359..855R,2012ApJ...756...16P}.

\section{Analysis}

Let us consider the merger of two NS producing a BH of radius $r_{BH}=3.3 \times 10^5$ cm. The $e^{\pm}$ plasma, of density $n_e=10^{33}$ part/cm$^3$, is generated in a shell of thickness $\Delta R$ around the event horizon. After annihilation to photons and neutrinos, a small fraction $n_{e,0}$ still remains as pairs. Due to the radiation pressure the shell expands at constant width $\Delta R$, making the volume $V$ increase and the density $n_e$ decrease until the transparency condition is reached. 
For $n_{e,0}=10^{26}$ part/cm$^3$ and $\Delta R=10^5$ cm, the number of particles in the shell is given by

\begin{equation}
N=n_{e,0} V=n_{e,0} \, \frac{4}{3} \pi [(R_{BH}+\Delta R)^3 - R_{BH}^3] = 1.82 \times 10^{43}.    
\end{equation}

\begin{figure}
\centering
\includegraphics[width=0.7\hsize]{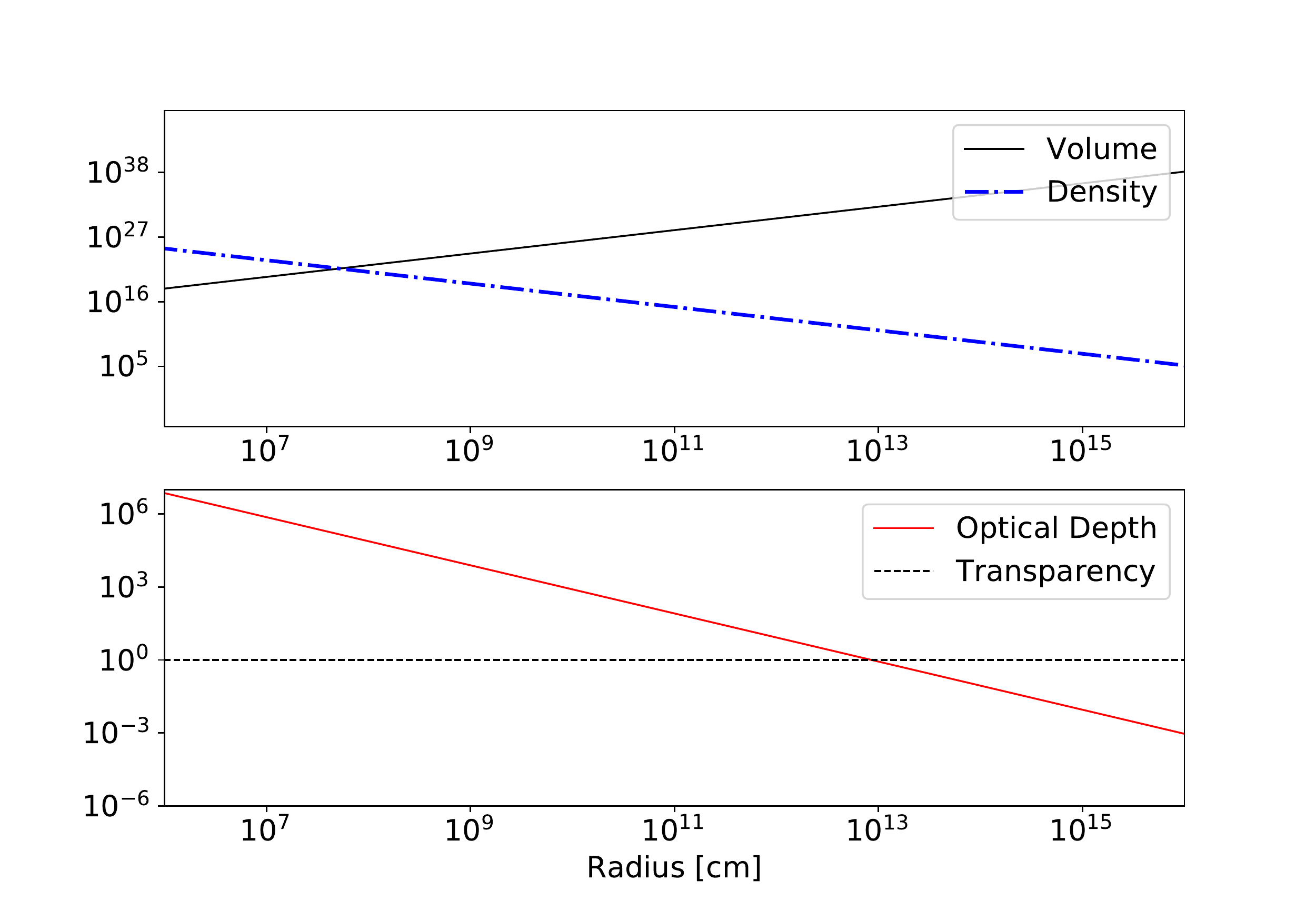}
\caption{Volume of the $e^{\pm}$ plasma-shell, density and optical depth as a function of the radius. The shell expands at constant width due to radiation pressure until it becomes optically thin. The transparency condition is given by $\tau (R) \leq 1$ (region below the dotted line).} 
\label{fig:tau}
\end{figure}

This number will remain constant during the expansion. The optical depth is given by

\begin{equation}
\tau(R)= \sigma_T n_e(R) \, R,
\end{equation}
where $\sigma_T= 6.652 \times 10^{-25}$ cm$^2$ is the Thomson cross section \cite{2006RPPh...69.2259M}. The optical depth will decrease as the shell expands until it falls to values below unity (transparency condition). At this moment photons are released and can travel towards the observer on Earth. Figure \ref{fig:tau} shows the behavior of the density, volume and optical depth as a function of radius. For the parameters just considered, the transparency condition is reached at the transparency radius $R=R_{tr}=9.65 \times 10^{12}$ cm.

\begin{figure}
\centering
\includegraphics[width=0.4\hsize]{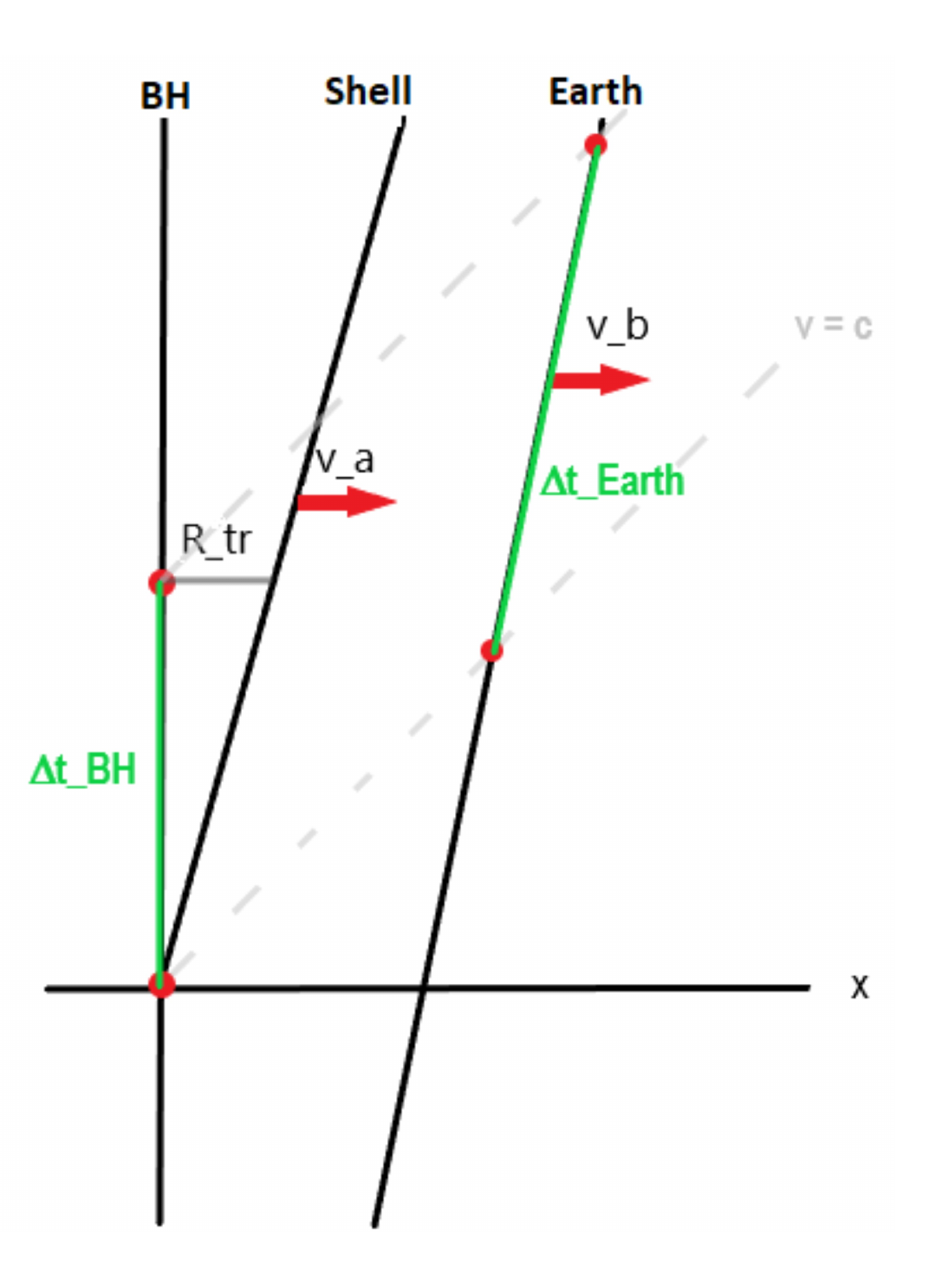}
\caption{Spacetime diagram illustrating the Lorentz transformation between three systems: the first one (BH) is the system in which the BH is at rest, the second one (Shell) is the system in which the shell is at rest, with relative velocity $v_a$ with respect to the BH. The third one is the observer frame on Earth, which is receding from the BH at a speed $v_b=H_0 D$. Here $H_0$ is the Hubble constant and $D$ is the distance between the BH and the observer.} 
\label{fig:Lorentz}
\end{figure}

Now that we have calculated the transparency radius we need to know how much time has elapsed since the moment neutrinos were created. In a frame for which the BH is at rest, an observer will see the shell travel away from him at $\Gamma \sim 100$, that is, at a velocity $v_a \sim 0.999949 c$. This gives the time interval $\Delta t_{BH}$,
\begin{equation}
\Delta t_{BH}= (R_{tr}-R_{BH})/v_a = 323.01 \, \rm{s}.
\end{equation}
We need to express $\Delta t_{BH}$ as seen by an observer on Earth. We assume that the BH is at the center of the galaxy, which is receding from the Milky Way at a velocity 
\begin{equation}
v_b= H_0 D,
\end{equation}
with $H_0= 70^{+12}_{-8}$ (km/s)/Mpc being the Hubble constant and $D$ the distance between the BH and the observer on Earth. Let us consider that the source is at a distance $D \sim 3240$ Mpc which, according to the Hubble law \cite{1929PNAS...15..168H}, corresponds to $z \sim 0.9$. This gives $v_b= 226800$ km/s, or, in units of $c$, $v_b= 0.756 c$. According to Lorentz transformations (see Fig. \ref{fig:Lorentz}), we obtain the time interval seen at 
Earth $\Delta t_{Earth}$, which is given by the expression
\begin{equation}
\Delta t_{Earth}=\sqrt{\frac{1+v_b/c}{1-v_b/c}} \, \Delta t_{BH} \approx 854.57 \, \rm{s}.
\end{equation}

This would be the delay between neutrinos and photons seen by an observer on Earth. 

\section{On the observability of the neutrino flux at Earth}
\textbf{In \cite{2019ApJ...872...73P} we have analysed the effects due to oscillations and decoherence on the neutrino flux accompanying a short GRB (NS-NS merger) at redshift $z=0.9$. We have considered $e^--e^+$ pair annihilation as the only source for neutrinos in order to simplify the model. According to the results presented in \cite{2011JPhCS.309a2029K}, the calculated neutrino flux at Earth is too small to be observed by the current generation of detectors, though the expected energy lies in the range of SN neutrinos \cite{Spiering}. However, this problem may be overcome by taking into account other channels for neutrino production \cite{Janka,Becerra} that could contribute to increase the total neutrino flux. Another way to increase the observability would be to lower the detector sensitivity threshold, as expected for future experiments. In any case, the increase of the flux will not modify the calculated time delay between neutrinos and photons arriving on Earth.}

\section{Conclusions}

We have estimated a delay of $\Delta t_{Earth} \approx 854.57$ s between the neutrino arrival at the detector on Earth and the gamma-ray detection. We have considered that $e^{\pm}$ pair-annihilation is the main channel for neutrino and photon production. We described the dynamics of the source as done in the framework of the Fireshell model. If this prediction were to be verified, there would be enough time available since the neutrino detection so as to alert other observatories, which would be able to follow the GRB emission from very early stages. 

\acknowledgments

This work has been partially supported by the National Research Council of Argentina (CONICET) by the grant PIP 616, and by the Agencia Nacional de Promoci\'on Cient\'ifica y Tecnol\'ogica (ANPCYT) PICT 140492. A.V.P and O.C. are members of the Scientific Research career of the CONICET.

\end{document}